\documentclass[amssymb,twocolumn,aps,pra]{revtex4-2}
\usepackage{soul}
\usepackage{xcolor}
\usepackage{graphicx}
\usepackage{amsmath}
\usepackage{dcolumn}
\usepackage{hyperref}
\usepackage{natbib}
\usepackage{textcase}
\usepackage{booktabs}
\usepackage[english]{babel}
\usepackage{mathtools}
\DeclarePairedDelimiter\bra{\langle}{\rvert}
\DeclarePairedDelimiter\ket{\lvert}{\rangle}
\DeclarePairedDelimiterX\braket[2]{\langle}{\rangle}{#1 \delimsize\vert #2}
\usepackage{amsmath}
\begin{document}

\title{Topologically Protected Edge States in One-Dimensional Quantum Walks}
\author{Emily Maxey}
\author{Jacob Mansfield}
\author{Beth Thacker}
\author{Wade DeGottardi}
\email{wdegotta@ttu.edu}
\affiliation{Department of Physics \& Astronomy, Texas Tech University, Lubbock, TX}
\date{\today}

\begin{abstract}

Topological insulators host protected boundary states that are robust to disorder, making them attractive for applications and motivating their realization in engineered systems. Discrete-time quantum walks, which have been implemented in a variety of experimental platforms, can exhibit topologically nontrivial phases and their associated boundary states.  Here, we introduce a class of topological quantum walks with variable step lengths that offer theoretical insights into the topological classification of such walks. For example, these walks can access higher winding numbers and support multiple edge states. They also provide context for the reduction of topological protection from a $\mathbb{Z}$- to $\mathbb{Z}_2$-valued invariant when time-reversal symmetry is broken. The topologically protected edge states are investigated using a transfer-matrix approach that describes their spatial profiles and spin structures. These predictions are in excellent agreement with numerical results and, where applicable, with the Jackiw--Rebbi zero-mode solution of the Dirac equation. Taken together, our analysis provides a roadmap for designing quantum walks with control over the number, spatial profile, and spin structure of their topological edge states.

\end{abstract}

\maketitle

\section{Introduction}

The topological classification of band structures has become a central theme in condensed matter physics. Topologically nontrivial materials support robust boundary states that give rise to striking physical phenomena. One prominent example is the quantum spin Hall effect (QSHE), which arises from helical edge states whose direction of motion is tied to their spin. This spin--momentum locking suppresses backscattering by nonmagnetic impurities~\cite{fu_topological_backscattering_2007,roushan_topological_backscattering_2009} and makes these states attractive for applications. These properties have motivated extensive efforts to realize topological band structures in engineered systems. Topologically nontrivial bands and their associated boundary modes have now been realized in a variety of settings, including cold-atom systems, synthetic photonic systems, metamaterials, and mechanical systems~\cite{karski_quantum_2009,crespi_anderson_2013,cai_topological_metamaterial_2025,wang_coriolis_2015}. Although the concept of a topological insulator originated in the study of electronic materials, band topology and its associated boundary states can fundamentally reshape even single-particle dynamics.

Periodically driven systems provide a particularly versatile setting for realizing topological phenomena. A prominent example is the discrete-time quantum walk (DTQW)~\cite{aharonov_quantum_1993,kitagawa_Floquet_Systems_2010}. The DTQW is inspired by the classical random walk, in which a particle moves one step to the right or left according to the outcome of a coin toss. In a quantum walk, an internal degree of freedom, which we refer to as spin, plays the role of the coin: a unitary spin rotation replaces the coin toss, and a spin-dependent translation moves the particle. Quantum walks have been studied in connection with coherent transport in chemical and biological systems, including photosynthesis~\cite{tang_photosynthesis_2024}; the modeling of chemical reaction networks~\cite{chia_coherent_2016}; and the implementation of quantum-computing protocols~\cite{qiang_quantum_2024,cao_quantum_2025}. They have also been realized experimentally in a variety of physical platforms~\cite{broome_discrete_2010,karski_quantum_2009,schmitz_quantum_2009,zahringer_realization_2010,schreiber_2d_2012,schreiber_photons_2010,crespi_anderson_2013}. Of central interest here, DTQWs can exhibit symmetry-protected topological phases~\cite{kitagawaExploring2010}. Topologically protected edge states arise at interfaces between regions with distinct topological invariants~\cite{kitagawaExploring2010,asboth_bulk--boundary_2013} and have been observed experimentally in a one-dimensional photonic quantum walk~\cite{kitagawa_observation_2012}.
While early studies of DTQWs largely focused on phases with topological indices equal to $0$ or $1$~\cite{kitagawaExploring2010,obuseUnveilingHidden2015}, more recent work has demonstrated that multistep quantum walks can realize phases with higher winding numbers and multiple edge states~\cite{Xiao2018,Jia2021}.

Here, taking inspiration from long-range hopping in topological insulators~\cite{kitaevTopologicalFermions2011},we introduce a class of quantum walks with variable step lengths whose topological properties are readily analyzed. We use these walks to investigate how the number, spatial profile, and spin structure of the resulting edge states depend on the parameters of the walk. Our central analytical tool is a transfer-matrix approach that predicts these properties which, in the continuum limit, agree with the Jackiw--Rebbi zero-mode solution of the Dirac equation. We verify these analytical predictions through numerical calculations and use real-time simulations to demonstrate how the edge modes influence the dynamics of a quantum walk. This framework provides a pathway for designing topological quantum walks with control over the number, spatial profile, and spin structure of their edge states.

Beyond enabling a detailed analysis of the edge states, these generalized walks provide a concrete setting in which to illustrate how symmetry controls their topological classification. In particular, we show explicitly how breaking time-reversal symmetry while preserving particle-hole symmetry changes the relevant invariant from $\mathbb{Z}$-valued to $\mathbb{Z}_2$-valued, consistent with the known classification~\cite{kitagawaExploring2010}. Quantum walks also possess distinctive features arising from their periodically driven nature. Their quasienergies($\lambda$) are defined modulo $2\pi$, and particle-hole-symmetric walks can consequently support protected modes at both $\lambda=0$ and $\lambda=\pi$~\cite{obuseUnveilingHidden2015}. Our work examines the implications of higher winding numbers and Floquet quasienergy structure on the properties of the resulting edge states.

This paper is organized as follows. In Sec.~\ref{sec:defs}, we define the generalized quantum walk considered in this paper. Section~\ref{sec:classification} reviews the classification of topological walks. It also presents our analysis of the reduction of topological protection from $\mathbb{Z} \to \mathbb{Z}_2$ for walks that respect particle-hole symmetry but violate time-reversal symmetry. In Sec.~\ref{Sec:Protected Edge Modes}, we present our transfer matrix approach and apply it to the analysis of topologically protected edge states. This section contains the key results of the paper. Finally, in Sec.~\ref{sec:outlook}, we summarize our results and discuss the outlook for future applications of topological quantum walks.

\section{Defining the Quantum Walks}
\label{sec:defs}

\begin{figure}
\includegraphics[width=0.9\linewidth]{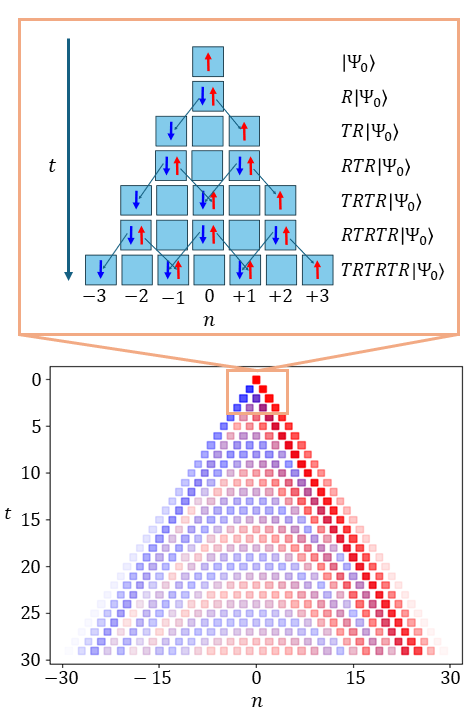}
\caption{Time evolution of the quantum walk generated by the unitary in
Eq.~(\ref{Eq:simpleUnitary}) with $\theta=\pi/2$. The probability of finding
the walker at lattice site $n$ after $t$ time steps is indicated by the 
intensity, while the local expectation value of the $z$ component of spin is
indicated by the color, where red (blue) indicates spin-up (down). The inset illustrates the evolution of the wave
function during the first few time steps.}
\label{fig:pyramid}
\end{figure}

In a classical random walk, a particle moves according to the outcome of a
coin toss. In a one-dimensional quantum walk, the role of the coin is played
by an internal two-level degree of freedom, which we refer to as spin. The
spin determines whether the particle moves to the left or to the right. The
state of the walker is therefore specified by both its position and spin:
$\ket{n}\otimes\ket{S}$ denotes a particle at lattice site $n$ in the spin
state $\ket{S}$.

Rather than being defined through a Hamiltonian, a discrete-time quantum walk
can be specified directly by the unitary operator governing a single time
step. A simple walk consists of a spin rotation followed by a spin-dependent
translation and is generated by repeated applications of
\begin{equation}
U=TR(\theta).
\label{Eq:simpleUnitary}
\end{equation}
Here,
\begin{equation}
R(\theta)=
\begin{pmatrix}
\cos(\theta/2) & -\sin(\theta/2) \\
\sin(\theta/2) & \cos(\theta/2)
\end{pmatrix}
\label{Eq:rotation}
\end{equation}
rotates the spin through an angle $\theta$ about the spin $y$ axis, while
\begin{equation}
T=\sum_n\left(
\ket{n+1}\bra{n}\otimes\ket{\uparrow}\bra{\uparrow}
+
\ket{n-1}\bra{n}\otimes\ket{\downarrow}\bra{\downarrow}
\right)
\label{Eq:translation}
\end{equation}
translates the particle conditionally according to its spin. A spin-up
particle moves one lattice site to the right, whereas a spin-down particle
moves one lattice site to the left. This coupling between the spin and motion
of the particle is reminiscent of spin--orbit coupling.

After $t$ complete time steps, the state of the walker is
\begin{equation}
\ket{\psi(t)}=U^t\ket{\psi(0)},
\end{equation}
where the integer $t$ counts the number of applications of $U$.
Figure~\ref{fig:pyramid} depicts the resulting dynamics. In contrast to the
diffusive spreading of a classical random walk, the quantum walk spreads
ballistically~\cite{aharonov_quantum_1993}: its propagating fronts move
outward with constant velocities.

To access a broader range of topological phases, we generalize the simple
quantum walk by allowing each spin component to move a variable number of
lattice sites during a translation. We define the multi-step translation
operator
\begin{equation}
T_{p,q}
=
\sum_n\left(
\ket{n+p}\bra{n}\otimes\ket{\uparrow}\bra{\uparrow}
+
\ket{n-q}\bra{n}\otimes\ket{\downarrow}\bra{\downarrow}
\right),
\label{eq:generalized-translation}
\end{equation}
where $p$ and $q$ are nonnegative integers. Under this operation, the spin-up
component moves $p$ lattice sites to the right, whereas the spin-down
component moves $q$ lattice sites to the left.

A single time step of the multi-step walk combines two spin rotations with
two spin-dependent translations and is described by the unitary operator
\begin{equation}
U_{(p_1,q_1,p_2,q_2)}
=
T_{p_2,q_2}R(\theta_2)T_{p_1,q_1}R(\theta_1).
\label{eq:unitary}
\end{equation}
Reading the operators from right to left, the spin is first rotated through
an angle $\theta_1$, followed by a translation generated by
$T_{p_1,q_1}$. A second rotation through $\theta_2$ is then followed by a
translation $T_{p_2,q_2}$. The conventional split-step quantum
walk~\cite{kitagawaExploring2010} is recovered for
$(p_1,q_1,p_2,q_2)=(1,0,0,1)$.

For a translationally invariant walk, the eigenstates of the unitary operator are plane waves of the form,
\begin{equation}
\ket{\Psi_k}
=
\sum_n e^{-ikn}\ket{n}\otimes\ket{\psi_k},
\label{eq:plane-wave}
\end{equation}
where $k$ is the pseudomomentum of the walk restricted to the first Brillouin zone $- \pi \leq k < \pi$ and $\ket{\psi_k}$ is a two-component spinor in the
$\{\ket{\uparrow},\ket{\downarrow}\}$ basis. Applying the unitary operator in
Eq.~(\ref{eq:unitary}) to $\ket{\Psi_k}$ preserves the plane-wave form while
transforming the spinor according to
$\ket{\psi_k}\rightarrow U_k\ket{\psi_k}$. The effective unitary is given by
\begin{equation}
U_k
=
\alpha_k\mathbb{I}
-i\boldsymbol{\eta}_k\cdot\boldsymbol{\sigma},
\label{eq:Uk}
\end{equation}
where $\boldsymbol{\sigma}=(\sigma_x,\sigma_y,\sigma_z)$. Explicit expressions
for $\alpha_k$ and $\boldsymbol{\eta}_k$ are given in the 
Appendix. Because $U_k$ is unitary, its eigenvalues are complex and may be written as
$e^{-i\lambda_{k,\nu}}$, where $\nu$ indexes the two bands. By defining an effective Hamiltonian $H_k$ through
$U_k=e^{-iH_k}$, we can recognize $\lambda_{k,\nu}$ as the quasienergies of our system. Quasienergies are defined modulo $2\pi$, so that
$\lambda_{k,\nu}$ and $\lambda_{k,\nu}+2\pi$ are identified as the same value.

We define the total rightward and leftward translation distances after one time step as
\begin{equation}\label{Eq: P and Q}
P=p_1+p_2,
\qquad
Q=q_1+q_2.
\end{equation}
For balanced walks with $P=Q$, the coefficients $\alpha_k$ and
$\boldsymbol{\eta}_k$ are real, and the eigenvalues of $U_k$ take the form
\begin{equation} \label{Eq: expanded Eigenvalue}
e^{\mp i\lambda_k}
= \alpha_k\mp i\lvert\boldsymbol{\eta}_k\rvert.
\end{equation}
These walks can possess gaps at quasienergies $\lambda=0$ and $\lambda=\pi$
for generic values of the rotation angles.

\begin{figure}
    \centering
    \includegraphics[width=\linewidth]{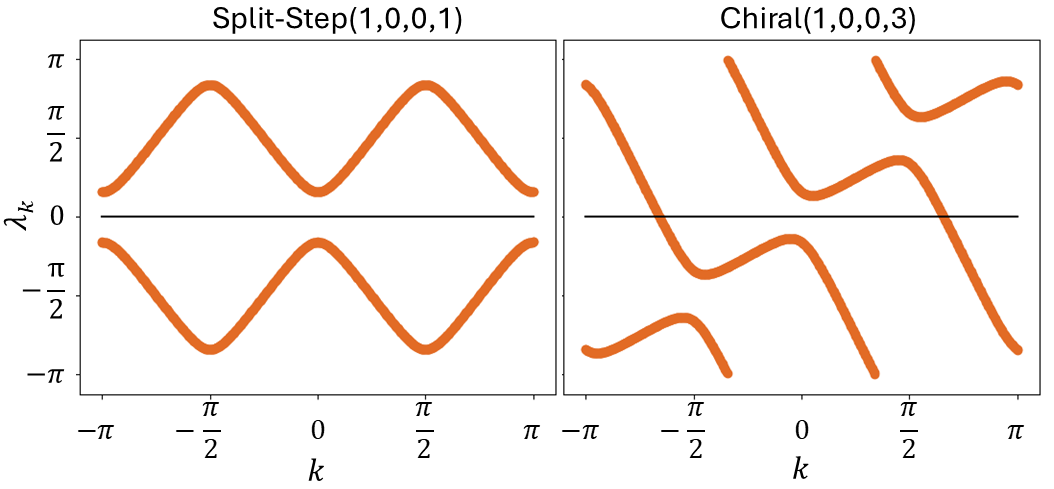}
    \caption{Quasienergy band structures for balanced and unbalanced
    multi-step quantum walks. (left) The balanced walk $(p_1,q_1,p_2,q_2) = (1,0,0,1)$. Any walk with $P=Q$ possesses
    quasienergy gaps for generic values of $\theta_1$ and $\theta_2$. On the other hand, any unbalanced walk, such as $(1,0,0,3)$ (right), does not possess gaps.}
    \label{fig:band_structure}
\end{figure}

For $P\neq Q$, the total rightward and leftward translation distances are
unequal. In this case, the quasienergy band winds around the two-torus defined
by $k$ and $\lambda_k$~\cite{kitagawa_Floquet_Systems_2010,
asboth_bulk--boundary_2013}. This winding precludes a global
quasienergy gap, as illustrated in Fig.~\ref{fig:band_structure}. Nevertheless, pseudogaps are present and can lead to edge states that, while not topologically protected, exhibit many of the same desired processes. These modes are discussed in Sec.~\ref{Sec:Protected Edge Modes}.

\section{Topological Classification of Quantum Walks}

\label{sec:classification}

The band structure associated with insulators can be classified according to its topological properties~\cite{chiu_classification_2016, teo_topological_2010}.This classification identifies $\mathbb{Z}$- or $\mathbb{Z}_2$-valued topological invariants that indicate the existence and number of the protected edge modes the system supports. Such invariants are only defined for gapped systems and can only change values if a gap closes and then reopens. All the gapped walks considered here respect particle-hole symmetry, and as will be demonstrated in Sec.~\ref{Sec:Protected Edge Modes}, the relevant gaps occur at $\lambda = 0$ and $\pi$. 

Following Ref.~\cite{kitagawaExploring2010}, we consider the symmetry operators $\mathcal{P}$ and $\mathcal{T}$, which are related to particle-hole and time-reversal symmetries, respectively. For convenience, here we work directly with the unitary $U_k$, defined above Eq.~(\ref{eq:Uk}), rather than the Hamiltonian. Particle-hole symmetry is formally defined as the existence of anti-unitary $\mathcal{P}$, such that $\mathcal{P}H_k\mathcal{P}^{-1}=-H_{-k}$\cite{teo_topological_2010}. In terms of the unitary $U_k$, the condition is
\begin{equation}
\mathcal{P} U_k \mathcal{P}^{-1} = U_{-k}.
\label{eq:particle-hole}
\end{equation}
Taking $\mathcal{P} = K$, where $K$ is the complex conjugation operator, we find that Eq.~(\ref{eq:particle-hole}) holds if and only if $P=Q$, a condition that is already required for the walks we consider to be gapped. This gives
\begin{equation}
    \boldsymbol{\eta}_k=\begin{pmatrix}
            c_2 s_1\sin (p_1+p_2)k - c_1s_2 \sin (q_1-p_2) k \\ 
            c_2s_1\cos(p_1+p_2)k + c_1 s_2 \cos(q_1-p_2)k \\ 
           -c_2 c_1 \sin (p_1+p_2)k - s_1 s_2\sin (q_1-p_2)k
    \end{pmatrix},
\label{eq:eta}
\end{equation}
where $c_i = \cos(\theta_i/2)$ and $s_i = \sin(\theta_i/2)$. 

\begin{figure}
    \centering
    \includegraphics[width=\linewidth]{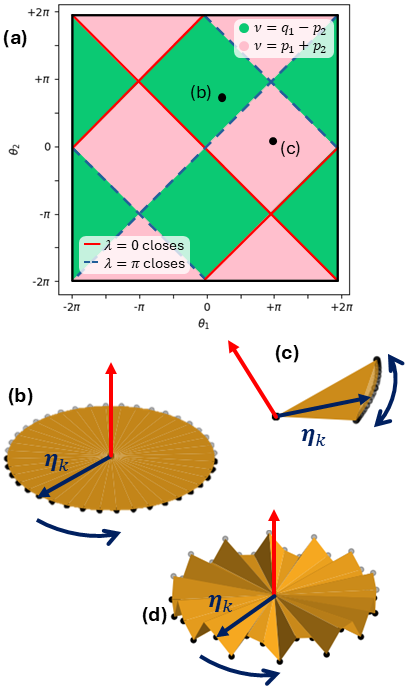}
    \caption{(a) Topological phase diagram indicating value of the winding number  $\nu$, which counts the number of times $\boldsymbol{\eta}_k$ winds for $-\pi \leq k < \pi$ as a function of $\theta_1$ and $\theta_2$. At the boundaries of the topologically distinct regions, the walk is gapless and thus $\boldsymbol{\eta}_k$ vanishes for one or more values of $k$. (b) Winding of $\boldsymbol{\eta}_k$ for a split-step walk $(p_1,q_1,p_2,q_2)=(1001)$ with $(\theta_1, \theta_2)=(\pi,\frac{\pi}{10})$. The chiral vector $\boldsymbol{A}$, indicated in red, is perpendicular to $\boldsymbol{\eta}_k$ for all $k$. (c) Plot of $\boldsymbol{\eta}_k$ for the split step walk, with $(\theta_1, \theta_2)=(\frac{\pi}{4},\frac{3\pi}{4})$ corresponding to a winding of $0$. In this case, $\boldsymbol{\eta}_k$ does not fully wind around the origin. (d) Value of $\boldsymbol{\eta}_k$ for a $(p_1,q_1,p_2,q_2)=(2,2,1,1)$, non-$\mathcal{T}$-symmetric, walk with $(\theta_1, \theta_2)=(\pi,\frac{\pi}{10})$.}
    \label{fig:phase-diagram}
\end{figure}

Time-reversal symmetry $\mathcal{T}$ plays a key role in the topological classification of the walks considered here. Derived from its definition, $\mathcal{T}H_k\mathcal{T}^{-1}=H_{-k}$ \cite{teo_topological_2010}, time-reversal symmetric walks must satisfy
\begin{equation}
\mathcal{T} U_k \mathcal{T}^{-1} = U^\dagger_{-k},
\label{eq:time-reversal}
\end{equation}
for anti-unitary $\mathcal{T}$.  Time-reversal symmetry can be written in terms of a particle-hole and a chiral symmetry operator $\Gamma$, in the form $\mathcal{T} = \Gamma \mathcal{P}$. A particle-hole symmetric walk that possesses chiral symmetry
\begin{equation}\label{eq:chiral_operator}
    \Gamma U_k\Gamma^{-1}=U_k^\dagger
\end{equation}
 with unitary $\Gamma$ also satisfies Eq.~(\ref{eq:time-reversal}). Given that all our considered walks respect particle-hole symmetry, the classification of the walks is thus controlled by chiral symmetry $\Gamma$.

We take the chiral symmetry operator to have the form $\Gamma=e^{-i\frac{\pi \boldsymbol{A}\cdot \boldsymbol{\sigma}}{2}}$, where $\boldsymbol{A}$ is a constant vector independent of $k$~\cite{kitagawaExploring2010}. For a unitary $U_k$ of the form of Eq.~(\ref{eq:Uk}), we find that
\begin{equation}
    \Gamma U_k\Gamma^{-1} = U_k^\dagger-2i(\boldsymbol{A}\cdot \boldsymbol{\eta}_k)(\boldsymbol{A}\cdot \boldsymbol{\sigma}).
\end{equation}
Hence, Eq.~\ref{eq:chiral_operator} is only satisfied provided that $\boldsymbol{A}\cdot\boldsymbol{\eta}=0$ for all $k$~\cite{kitagawaExploring2010}. For the walks considered here, chiral symmetry is satisfied for $p_1=q_2$ or equivalently $q_1=p_2$. In this case, $\boldsymbol{A} = (s_1, 0, c_1)$. However, when $q_1 \neq p_2$, the $\sin(q_1-p_2)k$ terms in Eq.~(\ref{eq:eta}) no longer vanish. These terms are responsible for the wobbling of the vector $\boldsymbol{\eta}$ as $k$ varies from $-\pi$ to $\pi$ that is evident in Fig.~\ref{fig:phase-diagram}d. This wobbling precludes the existence of a vector $\boldsymbol{A}$ independent of $k$, and thus chiral symmetry is violated. We note that this definition of time-reversal symmetry is a generalization of the standard definition~\cite{teo_topological_2010}, which is recovered if we set $\boldsymbol{A} = (0,1,0)$.  A summary of the symmetry properties and associated Altland Zirnbauer topological classification is given in Table~I.

\begin{table}\label{table:classifications}
    \begin{tabular}{ |c|c|c|c| } 
        \hline
        Conditions & $\mathcal{P}^2$ & $\mathcal{T}^2$ & Classification\\
        \hline
         $P=Q$ & $1$ & X & $\mathbb{Z}_2$ \\
         $p_1-q_2=0$ & $1$ & $1$ & $\mathbb{Z}$\\ 
    \hline
    
    \end{tabular}
    \caption{Symmetries of general walks. 
     For various step length conditions, we present the derived symmetries and the corresponding topological classification as given in \cite{altland_nonstandard_1997, zirnbauer_riemannian_1996, ryu_topological_2010} }
\end{table}
Walks that respect time-reversal ($\mathcal{T}$) symmetry have topological phases characterized by an integer-valued ($\mathbb{Z}$) topological invariant, which counts the number of times that $\boldsymbol{\eta}_k$ winds in the plane defined by $\mathbf{A}$ as $k$ goes from $-\pi$ to $\pi$. The form of $\boldsymbol{\eta}_k$ given in Eq.~(\ref{eq:eta}) makes it clear that this winding number is either $p_1 + p_2$ or $q_1 - p_2=0$, depending on the relative magnitude of $c_2 s_1$ and $c_1 s_2$. For the marginal case $| c_2 s_1 | = | c_1 s_2 |$, a gap at $\lambda = 0$ or $\pi$ closes and the invariant is undefined. Figure~\ref{fig:phase-diagram}a shows the topological phase diagram for a range of $\theta_1$ and $\theta_2$. 

The topological invariant counts the number of protected modes at the edge of a topological system, thus this result indicates that there would be $P=p_1+p_2$ such modes, which follows directly from the definition of the walks considered here. Consider the action of $U$ on a walker initially located at the origin. After a single application of the unitary (\ref{eq:unitary}), there is non-zero amplitude for the walker to be on sites $n = P, p_1-q_2,-q_1+p_2$ and $-Q$. For a walk with $\mathcal{P}$-and $\mathcal{T}$-symmetries, we have $p_1=q_2$ and $P=Q$ and thus the possible sites are $n = \pm P$ and the origin $n = 0$. Repeated applications of the unitary will similarly show that the walker only exists on the sublattice $n = mP$, where $m$ is any integer. In fact, this same reasoning shows why there are $P$ distinct edge modes: each mode corresponds to starting a walker in any one of the $P$ distinct sublattices. 

We now consider the nature of the topological invariant for walks that respect $\mathcal{P}$ symmetry but violate $\mathcal{T}$ symmetry, which have $q_1 \neq p_2$. In order to investigate the nature of the topological invariant in this case, we consider the phase boundary $\theta_1 = \theta_2$ and track the closure of the bulk gaps. As discussed above, a topological invariant can only change if a bulk gap closes, either at $\lambda = 0$ or $\pi$. From Eq. (\ref{Eq: expanded Eigenvalue}), this requires $|\boldsymbol{\eta}_k|$ to vanish for some $k$. For $k=\pi$,
\begin{equation}
    \boldsymbol{\eta}_\pi(\theta_1=\theta_2)=
    \begin{pmatrix}
        0 \\
        c_1 s_1[(-1)^{p_1+p_2}+(-1)^{q_1-p_2}] \\
        0
    \end{pmatrix},
\label{eq:bulk_gap}
\end{equation}
and we see that whether a bulk gap closes is dependent on the relative parities of $p_1 + p_2$ and $q_1 - p_2$. This demonstrates that the topological invariant in this case is not $\mathbb{Z}$-valued, but is $\mathbb{Z}_2$-valued, consistent with the standard classification Ref.~\cite{kitagawaExploring2010}. On the other hand, if $p_1+p_2$ mod $2 = q_1-p_2$ mod $2$, it is possible to cross the $\theta_1 = \theta_2$ phase boundary without closing a bulk gap. In fact, along this line, the bulk gap only closes at isolated points. In contrast, both the $\lambda = 0$ and $\lambda = \pi$ gaps close as the $\theta_1 = -\theta_2$ boundary are crossed, which is related to the fact that the topology of these walks cannot be fully captured by a single invariant.

\section{Topologically Protected Edge Modes}
\label{Sec:Protected Edge Modes}

\begin{figure*}
    \centering
    \includegraphics[width=\textwidth]{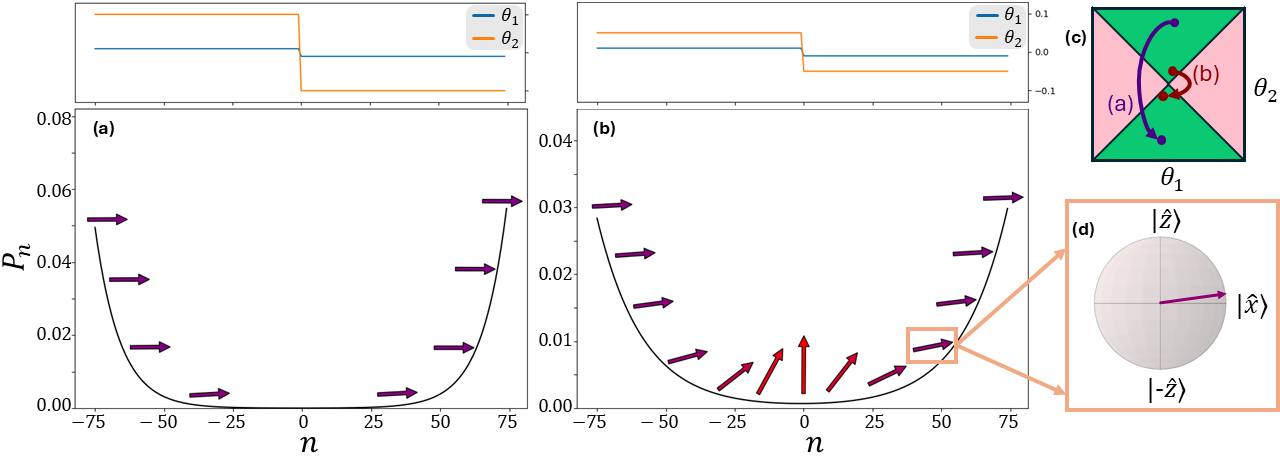}
    \caption{Edge state associated with $\lambda = 0$ for a split-step walk $(p_1,q_1,p_2,q_2)=(1,0,0,1)$. (a) $\lambda = 0$ edge state localized near $n = \pm 75$ at which both $\theta_1$ and $\theta_2$ change sign. The arrows represent the spin-states at various sites. (b) Edge states with smaller changes in $\theta_1, \theta_2$ across the boundaries lead to longer localization lengths. At regions where both eigenvectors of the transfer matrix control the spin, spatially dependent spin structure arises. (c) Topological phase diagram given in Fig.~\ref{fig:phase-diagram}a indicating the angles used in (a) and (b). For a description of these phases in terms of a single topological invariant, these phases are the same. However, as discussed in the text, due to the fact that there are gap closures at both $\lambda = 0$ and $\lambda = \pi$, a full description of the topology requires introducing a second topological invariant~\cite{obuseUnveilingHidden2015}.}(d) Example spin shown in the $xz$-plane. These arrows follow the same color convention as Fig \ref{fig:pyramid}, wherein red (blue) signifies spin up (down). 
    \label{fig:edge_modes}
\end{figure*}

As discussed in the previous section, the walks considered here respect particle-hole symmetry $\mathcal{P}$. In standard (continuous time) topological insulators, protected edge states are pinned at zero energy, provided that finite-size hybridization effects are neglected. The reason for this is that particle-hole symmetry dictates that for an eigenstate $|\psi\rangle$ with energy $E$, there is an associated state $\mathcal{P} | \psi \rangle$ with energy $-E$. There is nothing to prevent such states from being perturbed and moved out of (or into) the gap. On the other hand, a single state at $E = 0$ is pinned there by the requirement of particle-hole symmetry. There is no way to remove a zero-energy mode without closing a gap. On the other hand, discrete time quantum walks are described by a quasi-energy spectrum; this allows for considerably richer behavior. In particular, states with quasi-energies $\lambda = 0$ or $\pi$ are their own particle-hole conjugates and thus pinned at these quasi-energies. A full description of the topology of these states requires two topological invariants in order to track gap closures at $\lambda = 0$ and $\pi$~\cite{obuseUnveilingHidden2015}. For example, the topological phase diagram in Fig.~\ref{fig:phase-diagram}a indicates whether a given phase boundary is associated with the closure of a gap at $\lambda = 0$ or $\lambda = \pi$. 

While topological invariants indicate the existence of edge modes, a complete description of their physics requires solving the equations of motion for these modes, which are the eigenvalue equation $U |\lambda\rangle = e^{i \lambda} |\lambda\rangle$, for $\lambda = 0$ or $\pi$. Here, as will become clear from our derivation of the transfer matrix, our analysis applies to walks that obey both $\mathcal{P}$ and $\mathcal{T}$ symmetries. Since we are ultimately interested in the spatial and spin structure of these modes, we introduce
\begin{eqnarray}
  \langle \uparrow, n | \lambda \rangle &=& \alpha_n,  \\
  \langle \downarrow, n | \lambda \rangle &=&  \beta_n,
\label{eq:wavefunction}
\end{eqnarray}
where the complex functions $\alpha_n$ and $\beta_n$ are the amplitude for the walker to be at position $n$ with spin $| \uparrow \, \rangle$ and $| \downarrow \, \rangle$, respectively. The probability of finding the particle at a site $n$ is $P_n = |\alpha_n|^2+|\beta_n|^2$.

A transfer matrix may be derived from the eigenvalue equation
\begin{equation}
U \ket{\Psi_k} = e^{- i \lambda} \ket{\Psi_k}.
\label{eq:eigenvalue}
\end{equation}
According to the argument made in Sec.~\ref{sec:classification}, for $\mathcal{P}$- and $\mathcal{T}$-symmetric walks, the application of the unitary $U$ (\ref{eq:unitary}) to a walker initially at the origin leads to amplitude for the walker to be at sites $n = -P$, the origin, and $n = P$. Thus, the eigenvalue equation~(\ref{eq:eigenvalue}) leads to relations between wave functions on adjacent sites of the sublattice $n = P m$, where $m$ is any integer. The transfer matrix is obtained by solving for the amplitudes $\alpha_n$ and $\beta_n$ at $n = P$ in terms of their values at $n = 0$ and $n = -P$. However, by translational invariance, the resulting equations are valid for all $n$.  We thus obtain
\begin{align}
&
\begin{pmatrix}
    \alpha_{n+P}  \\
    \beta_{n+P}  
\end{pmatrix} 
 = \nonumber \\ 
& \begin{pmatrix}
     \pm c_1/c_2 & -(s_2 \pm s_1)/c_2 \\
    -(s_2\pm s_1)/c_2 & (2s_1s_2\pm 1 \pm s_1^2)/(c_1c_2)  
\end{pmatrix}
\begin{pmatrix}
    \alpha_{n}  \\
    \beta_{n}  
\end{pmatrix},
\label{eq:transfer_matrix}
\end{align}
where the $+$ sign corresponds to a $\lambda = 0$ mode and the $-$ sign to $\lambda = \pi$ modes. The transfer matrix is expressed in the $\{ \ket{\uparrow}, \ket{\downarrow} \}$ basis and connects wave functions at sites that are separated by a distance $P$.

The transfer matrix in Eq.~(\ref{eq:transfer_matrix}) encodes the essential properties of an edge mode: its eigenvalues determine the spatial decay of the wave function, while its eigenvectors determine the spin structure. To illustrate this, consider the topologically protected edge mode shown in Fig.~\ref{fig:edge_modes}a, localized near $n=\pm75$. For this mode, $\theta_1\approx0$ and $\theta_2(n)=-\delta\theta \, \mathrm{sgn}(n)$, with $|\delta\theta| \ll 1$. According to the topological classification given in Sec.~\ref{sec:classification}, the walk is in the same topological phase for $n < 0$ and $n > 0$. However, as mentioned above, these phases are topologically distinct when a second topological invariant is introduced. This is related to the fact that there are two protected modes near $n = \pm 75$, one associated with $\lambda = 0$ and the other $\lambda = \pi$. Here, we focus on the $\lambda=0$-mode. For $n < 0$, the transfer matrix is, to linear order in $\delta\theta$,
\begin{equation}
\mathcal{M}\approx
\begin{pmatrix}
1 & -\delta\theta/2 \\
-\delta\theta/2 & 1
\end{pmatrix}.
\label{eq:approx_transfer_matrix}
\end{equation}
The system has periodic boundary conditions, so that the sites $n=75$ and $n=-75$ are identified and represent a single interface. Consequently, the two sides of the edge mode must be analyzed separately. In a gapped phase, one of the eigenvalues of the transfer matrix is greater than 1 in magnitude, while the other is less than one. For $n<0$, moving away from the interface and into the bulk corresponds to increasing $n$, so the physical edge mode is associated with the eigenvalue whose magnitude is less than one.

Since the transfer matrix (\ref{eq:approx_transfer_matrix}) is essentially the sum of the identity matrix and $(-\delta \theta/2)\sigma_x$, its eigenvectors are $+\hat{x}$ and $-\hat{x}$ spinors, with eigenvalues $1-\delta\theta/2$ and $1+\delta\theta/2$, respectively. The requirement that the wave function decays into the bulk therefore selects the $+\hat{x}$ spinor for $n<0$, consistent with the numerical results shown in Fig.~\ref{fig:edge_modes}a. Thus, the same transfer matrix simultaneously determines both the spin polarization and the spatial decay of the edge state. The corresponding probability density has a characteristic decay length of approximately $1/|\delta\theta|\approx10$, in agreement with the numerically obtained mode shown in Fig.~\ref{fig:edge_modes}a. Because this decay length is much smaller than the distance between the interfaces, the edge state is well localized and is dominated on each side by a single eigenvector of the transfer matrix, which explains why its spin polarization is approximately independent of position.

On the other hand, spatially-dependent spin structures arise when distinct eigenmodes spatially overlap, as is the case in Fig.~\ref{fig:edge_modes}b. For this mode, the localization length is larger because the magnitudes of $\theta_1$ and $\theta_2$ have been reduced. Another interesting feature of this mode is that half way between the left- and right-hand sides of the mode, the spin is fully polarized in the $+\hat{z}$-direction. This feature arises from the fact that $\theta_1$ and $\theta_2$ vanish at this point. From Eq.~(\ref{eq:transfer_matrix}), we see that the off-diagonal matrix elements of the transfer matrix vanish at this point and so the eigenvalues correspond to a $\hat{z}$-polarized state.

We remark on an interesting difference between edge states associated with the closure of gaps at $\lambda = 0$ and those with closures at $\lambda = \pi$ (this feature of the phase boundaries is indicated by solid and dashed lines in Fig.~\ref{fig:phase-diagram}a). Importantly, eigenvalues of the transfer matrix are positive for $\lambda = 0$ and negative for $\lambda = \pi$. Because these eigenvalues relate a wave function, i.e., the values of $\alpha_n$ or $\beta_n$, at neighboring sites, the $\lambda = 0$ edge modes have a uniform sign structure while $\lambda = \pi$ modes have a staggered sign structure. This is consistent with the fact that the sign structure of a $\lambda = 0$-mode does not change from one application of $U$ to the next, whereas a $\lambda = \pi$ mode alternates in sign. 

\begin{figure}
    \centering
    \includegraphics[width = \linewidth]{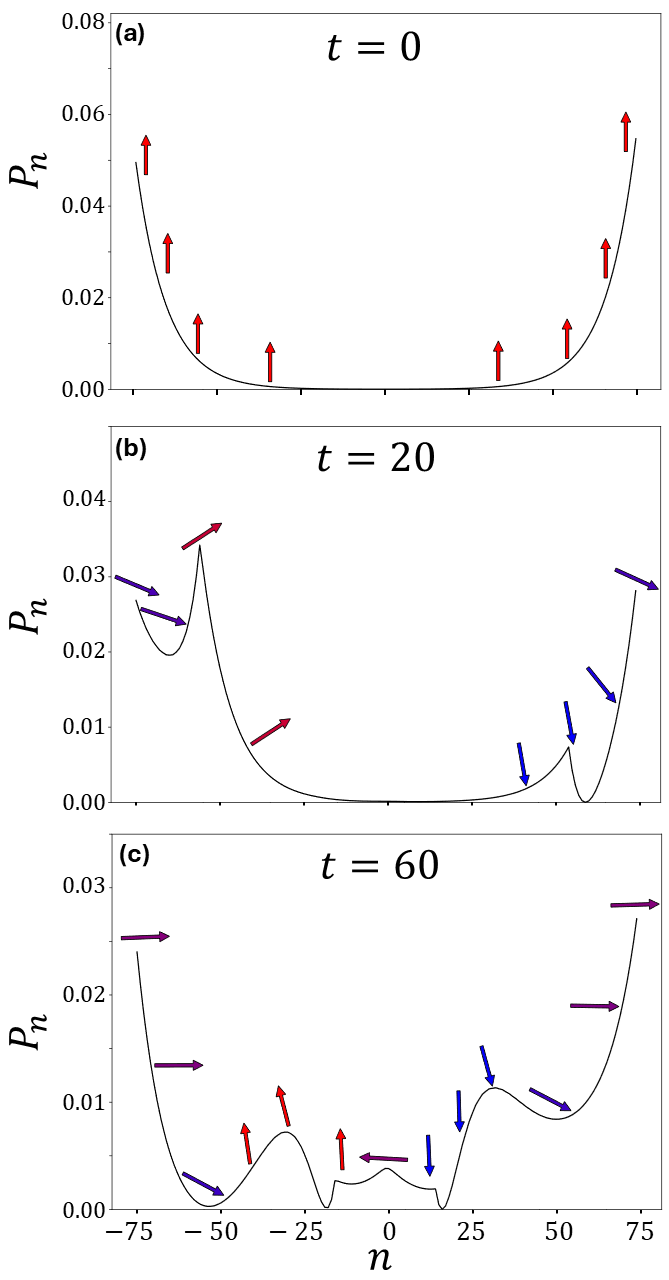}
     \caption{Edge state dynamics. Time evolution of a quantum walk $(p_1,q_1,p_2,q_2)=(1,0,0,1)$ with the same position-dependent coin as Fig. \ref{fig:edge_modes}a. (a) At $t=0$, the walk is initialized with the same spatial structure as the eigenmode in \ref{fig:edge_modes}a, but with a uniform spin in the $+\hat{z}$ direction. The wave function after (b) $t = 20$ and (c) $t = 60$ time steps. (c) the $+\hat{x}$ spin structure of the topologically-protected edge state begins to appear at $n \approx \pm 75$,.}
    \label{fig:dynamics}
\end{figure}

So far, we have focused on individual topological modes. Ultimately, the effects of these modes must be probed through the dynamics of a walker. Figure~\ref{fig:dynamics} gives a dramatic example of the effects of the mode. A walker is prepared in a state that matches the spatial structure of the topologically protected edge mode, but not its spin. Instead, the spin is polarized in the $+\hat{z}$-direction. In Fig.~\ref{fig:dynamics}, we see that the walker's dynamics ``purifies'' the edge state. Since only the $+\hat{x}$-spin state is stationary, this part of the initial wave function is left behind as the rest of the walk moves away from the edge.

It is instructive to describe the protected modes in the continuum limit. Here, we focus on $\lambda = 0$ modes and specialize to the case that $\theta_2 = 0$. In this case,
\begin{equation}
U_k = \left(
\begin{matrix}
      e^{-i P k} \cos \left(\theta_1/2\right)  & -e^{-i P k} \sin \left( \theta_1/2 \right)  \\
      e^{i Q k} \sin \left( \theta_1/2 \right)        &  e^{i Q k} \cos \left( \theta_1/2 \right)  \\
    \end{matrix}
\right),
\end{equation}
where $P$ and $Q$ are defined in Eq. \ref{Eq: P and Q}. We further take $\theta_1(n)$ to be a slowly varying function of $n$. Expanding $\mathcal{U}_k$ to first order in $\theta_1(x)$ and $k$, we obtain $\mathcal{U}_k \approx 1-iH$ where the Hamiltonian is
\begin{equation}
H =
\left(
\begin{matrix}
      -i P \partial_x   & - i \theta_1/2  \\
      i \theta_1/2       & 
      i Q \partial_x  \\
    \end{matrix}
\right).
\label{eq:chiral_ham}
\end{equation}
This Hamiltonian has been expressed in real space by taking $k \to - i \partial_x$.

Now, consider a region of space in which $\theta_1(x)$ is monotonically decreasing, as is the case in Fig.~\ref{fig:edge_modes}a. The Hamiltonian (\ref{eq:chiral_ham}) has a zero mode localized near the point $\theta(x) = 0$ of the form
\begin{equation}
\psi(x) = 
\left(
\begin{matrix}
 \sqrt{Q} \\
\pm\sqrt{P} 
\end{matrix}
\right) \exp \left( \pm \frac{1}{2\sqrt{P Q}} \int_0^x \theta_1(x') dx' \right),
\label{eq:Jackiw-Rebbi}
\end{equation}
where the upper and lower signs correspond to the two eigenmodes of $H$. This is the celebrated Jackiw-Rebbi solution of the Dirac equation. We observe that for the walk considered in Figure~\ref{fig:edge_modes}a, which has $P = 1$ and $Q = 1$, the $+$ sign recovers the behavior seen above: The spin points in the $+\hat{x}$-direction and the edge mode decays into the bulk.

An interesting extension of this solution is to consider the cases for which $P \neq Q$. In this case, the band structure is no longer insulating. Rather, it is chiral, and its gaplessness is protected by the fact that it has non-trivial winding on the surface of the $(k,\lambda)$ two-torus. Nevertheless, this band structure does admit a ``pseudogap'': in a limit range of $k$, there is local gap that is proportional to $\theta$. Investigating the robustness of these modes to disorder offers an interesting extension of the present work that extends the Jackiw-Rebbi solution to the case in which there are different speeds of light for right- and left-movers. 

\section{Summary and Outlook}
\label{sec:outlook}

In this work, we have performed a detailed analysis of the topological edge states of a generalized quantum walk with multi-step hopping. These generalized walks provide insight into their topological classification, and particularly into the reduction of topological protection from $\mathbb{Z}$ to $\mathbb{Z}_2$ that occurs when $\mathcal{T}$ symmetry is broken while $\mathcal{P}$ symmetry is preserved. The key result of our paper is the introduction of a transfer-matrix approach and its application to the analysis of topological edge states. This approach provides a detailed understanding of the spatial and spin structure of these edge states and reveals how these properties influence the dynamics of the quantum walk. Given that applications of topological quantum walks will likely rely on topologically protected edge modes, understanding how their properties are controlled by the unitary evolution of the walk is essential. The analysis and techniques introduced here, particularly the transfer-matrix approach, provide a first step toward designing and controlling these modes for future applications.

The authors wish to thank Kendra Jean-Jacques for her feedback and time in constructive conversation. We dedicate this work to the memory of our colleague and friend Dr. Ismael de Farias. We gratefully acknowledge his invaluable insights during the initial stages of this study.

\bibliography{refs}

\clearpage
\onecolumngrid
\appendix
\pagenumbering{alph}

\section{Supplementary Information}
\label{Sec: derivingUk}

We define $U_k$ as the action of the unitary on a single $k$-space.  The corresponding translation operator in the spin basis then becomes,
\begin{equation}
T_{p,q}=
    \begin{pmatrix}
        e^{-ikp} & 0 \\
        0 & e^{ikq}
    \end{pmatrix}.
\end{equation}
and the rotation operator is defined in Eq \ref{Eq:rotation}.  Thus we find

\begin{equation}
U_{p_1q_1p_2q_2}=T_{p_2q_2}R(\theta_2)T_{p_1q_1}R(\theta_1)=\begin{pmatrix}
        e^{-ik(p_1+p_2)}(c_1c_2-e^{ik(p_1+q_1)}s_1s_2) & -e^{-ik(p_1+p_2)}(c_2s_1 + c_1s_2e^{ik(p_1+q_1)}) \\ 
        e^{-ik(p_1-q_2)}(e^{ik(p_1+q_1)}c_2s_1+c_1s_2) & e^{-ik(p_1-q_2)}(c_1c_2e^{ik(p_1+q_1)}-s_1s_2)
    \end{pmatrix},
\end{equation}
where $c_i$($s_i$) is simply $\cos(\theta_i/2)$($\sin(\theta_i/2)$). Casting $U_{p_1q_1p_2q_2}$ in the form of Eq.~(\ref{eq:Uk}), we find that 
\begin{equation}\label{Eq:etaReal}
    \mathbb{R}e(\boldsymbol{\eta})=\frac{1}{2}\begin{pmatrix}
        c_2s_1[\sin(k(q_1+q_2))+\sin(k(p_1+p_2))]-c_1s_2[\sin(k(p_1-q_2))+\sin(k(q_1-p_2))]\\
        c_2s_1[\cos(k(p_1+p_2))+\cos(k(q_1+q_2))]+c_1s_2[\cos(k(p_2-q_1))+\cos(k(p_1-q_2))] \\
        -c_1c_2[\sin(k(p_1+p_2))+\sin(k(q_1+q_2))]+ s_1s_2[\sin(k(q_2-p_1))+\sin{(k(p_2-q_1))}]
    \end{pmatrix}
\end{equation}
\begin{equation}\label{Eq:etaImag}
    \mathbb{I}m(\boldsymbol{\eta})=\frac{1}{2}\begin{pmatrix}
         c_2s_1[\cos(k(q_1+q_2))-cos(k(p_1+p_2))]+c_1s_2[\cos(k(p_1-q_2))-\cos(k(q_1-p_2))]\\
        c_2s_1[\sin(k(p_1+p_2))-sin(k(q_1+q_2))]+c_1s_2[\sin(k(p_1-q_2))-\sin(k(q_1-p_2))] \\
        c_1c_2[\cos(k(p_1+p_2))-\cos(k(q_1+q_2))]+  s_1s_2[\cos(k(p_1-q_2))-\cos{(k(q_1-p_2))}]
    \end{pmatrix}
\end{equation}
and $\alpha_k$ is
\begin{equation}
    \mathbb{R}e(\alpha_k) = \frac{1}{2}(c_1c_2[\cos(k(p_1+p_2)) + \cos(k(q_1+q_2))] - s_1s_2[\cos(k(q_1-p_2))+\cos(k(p_1-q_2))])
\end{equation}
\begin{equation}
\label{eq:alphaImag}
    \mathbb{I}m(\alpha_k) = \frac{1}{2}(c_1c_2[\sin(k(p_1+p_2)) - \sin(k(q_1+q_2))]-s_1s_2[\sin(k(p_1-q_2))-\sin(k(q_1-p_2))]).
\end{equation}
A simplified form of $\boldsymbol{\eta}_k$ is shown in Eq. \ref{eq:eta} for gapped walks. These walks obey $P = Q$ and have $\alpha_k, \boldsymbol{\eta}_k \in \mathbb{R}$.

\end{document}